# Analytic Solution for Perturbed Keplerian Motion Under Small Acceleration Using Averaging Theory

Giacomo Curzi[1] and Dario Modenini[2]

*University of Bologna, Forlì 47121, Italy.*

**A novel approach is developed for analytic orbit propagation based on asymptotic expansion with respect to a small perturbative acceleration. The method improves upon existing first order asymptotic expansions by leveraging on linear systems and averaging theories. The solution starts with the linearization of Gauss planetary equations with respect to both the small perturbation and the six orbital elements. Then, an approximate solution is obtained in terms of secular and short period components. The method is tested on a low-thrust maneuver scenario consisting of a Keplerian orbit perturbed by a constant tangential acceleration, for which a solution can be obtained in terms of elliptic integrals. Results show that the positional propagation error is about one order of magnitude smaller with respect to state-of-the-art methods. The position accuracy for a LEO orbit, apart from pathological cases, is typically in the range of tens of meters for a dimensionless tangential acceleration of $10^{-5}$ after 5 orbital periods propagation.**

## 1    Introduction

Electric propulsion systems have attracting features for space missions due to their high specific impulse and reliability. On the other hand, they are limited in the achievable thrust magnitude, so that orbital manoeuvres typically require continuous, long lasting thruster firings. Furthermore, they feature non-negligible thrust variability in terms of both bias (De Bruijn, et al., 2017) and random (short-term) fluctuations (Randall, et al., September 15-20, 2019).

As new satellites and constellations are getting into space equipped with electric propulsion (Willis & D'Amico, 2018), collision avoidance (COLA) operations will be affected by the usage of low-thrust engines (Muelhaupt, et al., 2019). Authors in (Petit, et al., 2019) demonstrates that COLA maneuvers with low-thrust in Low Earth Orbit (LEO) can be realistically performed between 4 and 8 orbits ahead of the conjunction, that means around 12 h ahead. Nevertheless, the above-mentioned engines features prevent a straightforward extension of current COLA practices since i) a low thrust maneuver requires a significant burn time differently from impulsive ones, encumbering the maneuver optimization process and ii) the thrust uncertainty contributes to the overall position uncertainty (De Bruijn, et al., 2017), thereby influencing the effectiveness of the maneuver itself in decreasing collision risk.

---

[1] PhD candidate, Department of Industrial Engineering.

[2] Assistant Professor, Department of Industrial Engineering.

In such a context, the availability of fast trajectory propagation methods would be highly beneficial. Even though valuable efforts have been put towards analytic models for fast maneuver optimization, currently established techniques for low thrust optimization involve numerical solutions (direct or indirect methods) (Morante, et al., 2021).

Several authors developed very useful analytical solutions for the orbital motion under low thrust, most notably Gao (Gao, 2007), Bombardelli et al (Bombardelli, et al., 2011), Zuiani et al (Zuiani, et al., 2012) Di Carlo (Di Carlo, et al., 2021) and Gonzalo et al (Gonzalo & Colombo, 2021). Although with different approximations, the approaches are quite similar for all the orbital elements apart from the time element, which undergoes different, ad-hoc treatment. Gao provided approximate solutions using classical orbital elements for computing near-optimal low-thrust Earth-orbit transfers using tangential, inertial or piecewise constant out-of-plane thrust, including also the effect of J2. To this end, he integrated the Gauss planetary equations along the fast angular variable (eccentric anomaly) for the first five orbital elements. The time law was instead approximated as purely Keplerian, i.e. neglecting the small perturbation presence. Bombardelli et al. (Bombardelli, et al., 2011) developed an analytical solution under the assumption of a small constant tangential thrust again integrating the orbital dynamics along the sole fast variable, i.e. the angular position. They adopted a set of three generalized nonsingular orbital parameters and introduced a Sundman transformation for the time equation. In that case, the time-of-flight is integrated using a full first-order solution in the perturbing acceleration. Zuiani et al. performed the same kind of integration but removing the tangentiality assumption of the thrust. Di Carlo (Di Carlo, et al., 2021) lately extended the work of Zuiani (Zuiani, et al., 2012) and (Zuiani & Vasile, 2015) to include atmospheric perturbation, solar radiation pressure, third body and higher order harmonics. In all these three works the time equation was first reduced to its Keplerian version, and then integrated assuming first order variations in the remaining orbital elements.

Gonzalo et al (Gonzalo & Colombo, 2021) adopted a similar approach for solving the time equation, by incorporating variations of the semi-major axis and eccentricity into the time equation expanded at first order in thrust.

The resulting integral was then solved expanding the analytical solutions of semi-major axis and eccentricity to 4$^{th}$ order for small eccentricity. Authors also proposed to correct the outcome computed through the integration of Kepler's equation by adding the rotation of the line of apses, i.e. $\Delta\omega$.

One of the firsts applicative examples of such models in a collision avoidance maneuver scenario can be found in (Gonzalo, et al., 2020). Considering a tangential thrust dynamic, the authors adopted the eccentric anomaly as independent angular variable and built an analysis tool for collision avoidance considering the possibility of a low-thrust. Specifically, they provide and make use of analytic solution for the first five orbital elements, but integrate the time equation numerically, since the resulting accuracy was not sufficient for COLA purposes.

A common simplification in the cited methods is that integration of the equations of motion is carried out keeping constant all orbital elements except for the "fast" time variable. In the long-time horizon, this is a crude approximation especially for the semi-major axis whose actual value slowly drifts away from the analytical solution. This, in turn, negatively affects especially the time law in which the semi major axis variation plays a crucial role, as recognized for instance by Gonzalo et al in (Gonzalo & Colombo, 2021).

In this work we propose an approach to analytic orbit propagation under low thrust which seeks to improve with respect to existing methods by i) allowing the variation of all orbital elements during integration by linearization of Gauss Planetary Equations, ii) including in the time law equation the first order effect of the small perturbative acceleration. The method is then applied to the variational equations of classical orbital elements under a purely tangential constant acceleration, and an approximate solution is developed for the resulting system of non-autonomous, non-stationary linear differential equations, by exploiting averaging theory. A preliminary version of this work appeared as an extended abstract at the II Global Virtual Workshop of the Stardust-R network on Space Traffic Management and Resilient Space Environment.

The paper is organized as follows. Section 2 set up the problem and reviews the set of equations to be integrated. In a dedicated subsection, also some basics of averaging theory useful for the solution derivation is outlined. In Section 3 the proposed solution is obtained and in Section 4 is tested against state-of-the-art methods and numerical integration. Finally, conclusions are drawn in Section 5.

## 2 Mathematical model

### 2.1 *Preliminaries and notation*

We denote (matrices) vectors in (capitol) bold face, with components (rows, columns) indicated with light face subscripts. Subscript $r$ denotes the first 5 components or the upper left 5x5 block, respectively for a vector or a matrix. Italic characters are used for the primitive of a function. A plain superscript applied to a vector or matrix indicates the order in the parameter $\varepsilon$ of a Taylor expansion near $\varepsilon = 0$.

The orbital position of a spacecraft is described trough a set of Keplerian orbital elements $\boldsymbol{\alpha}^* = [\boldsymbol{\alpha}_r \quad \nu]^T \in \mathbb{R}^6$, where $\boldsymbol{\alpha}_r$ is a vector representing the orbit shape and orientation, whereas $\nu$ is a generic scalar angular variable linking the position along the orbit to time.

The differential equations describing the evolution of the six Keplerian orbital elements as a function of a perturbation accelerations $\boldsymbol{a}_p$, mostly known as Gauss Planetary Equations, can be written in the functional form:

$$\frac{d\boldsymbol{\alpha}^*}{dt} = \boldsymbol{\Phi}(t, \boldsymbol{\alpha}^*, \boldsymbol{a}_p), \tag{1}$$

For the upcoming derivation, we will study the GPE adopting the eccentric anomaly as angular variable, $\nu = E$, and the set of classical orbital elements semi-major axis, eccentricity, inclination, longitude of ascending node and argument of pericenter, i.e. $\boldsymbol{\alpha}_r = [a, e, i, \Omega, \omega]$. A solution can then be expressed by replacing the 6$^\text{th}$ orbital elements with time and adopting $E$ as the independent variable, i.e.:

$$\frac{d\boldsymbol{\alpha}}{dE} = \boldsymbol{f}(E, \boldsymbol{\alpha}, \boldsymbol{a}_p), \tag{2}$$

where:

$$\boldsymbol{f}(E, \boldsymbol{\alpha}, \boldsymbol{a}_p) = \begin{bmatrix} \boldsymbol{\Phi}_r(t, \boldsymbol{\alpha}^*, \boldsymbol{a}_p)\left(\frac{dE}{dt}\right)^{-1} \\ \frac{dt}{dE} \end{bmatrix},$$

$$\boldsymbol{\alpha} = [\boldsymbol{\alpha}_r \quad t]^T. \tag{3}$$

References (Bombardelli, et al., 2011), (Gonzalo, et al., 2020) and (Gonzalo & Colombo, 2021) had a similar approach.

*2.2  First order asymptotic expansion of Gauss Planetary Equations*

We assume that the perturbing acceleration can be expressed as a function of a dimensionless parameter $\varepsilon$, i.e.

$$\frac{\boldsymbol{a}_p}{\mu/a_c^2} = \varepsilon \boldsymbol{g}(E, \boldsymbol{\alpha}), \tag{4}$$

with $\varepsilon \ll 1$ and $\|\boldsymbol{g}\|$ in the order of unity or less, thus enforcing the perturbation magnitude to be much smaller than that of the Keplerian acceleration. Our approach starts with approximating Eq.(2) to first order in $\varepsilon$ as:

$$\frac{d\boldsymbol{\alpha}}{dE} \approx \boldsymbol{f}^0(E, \boldsymbol{\alpha}) + \varepsilon \boldsymbol{f}^1(E, \boldsymbol{\alpha}, \boldsymbol{g}(E, \boldsymbol{\alpha})), \tag{5}$$

In Eq. (5), $\boldsymbol{f}^0$ represents the evolution of the unperturbed $\boldsymbol{\alpha}$ (i.e. a Keplerian orbit), thus its first 5 components are zero, while the last component embeds Kepler's equation for the assumed "time" variable. In our case:

$$f_6^0 = \left.\frac{dt}{dE}\right|_{\varepsilon=0} = \frac{n}{1 - e\cos E} \tag{6}$$

The advantage of this approach is that $\boldsymbol{f}^1(E, \boldsymbol{\alpha})$ can be often integrated in analytic form when evaluated at $\boldsymbol{\alpha} = \boldsymbol{\alpha}_c$, as demonstrated in (Gonzalo & Colombo, 2021), where $\boldsymbol{\alpha}_c$ is the chief (or initial) Keplerian orbital element vector.

### 2.3 Approximation of Gauss Planetary Equations under tangential low thrust

We assume as a reference application the one of a continuous low-thrust maneuver providing a small, constant acceleration, in tangential direction $a_\tau = \varepsilon \mu / a_c^2$. To work in a dimensionless setting, we adopt the chief (unperturbed) semi-major axis and reciprocal mean motion as the units of length and time, respectively $a_c$ and $1/n_c$, and define the corresponding dimensionless elements as:

$$\tilde{a} = \frac{a}{a_c}, \tilde{t} = n_c t. \tag{7}$$

In the remaining of the work, we will refer to a dimensionless $\boldsymbol{\alpha} = [\tilde{a}, e, i, \Omega, \omega, \tilde{t}]^T$. With these assumptions, GPEs for a constant tangential perturbation can be written as (see Battin (Battin, 1999)):

$$\frac{d}{d\tilde{t}}[\tilde{a}, e, i, \Omega, \omega]^T = \varepsilon \begin{bmatrix} 2\tilde{a}^{3/2}\sqrt{\frac{1 + e\cos E}{1 - e\cos E}} \\ 2\tilde{a}^{1/2}\frac{(1 - e^2)\cos E}{\sqrt{1 - e^2 \cos^2 E}} \\ 0 \\ 0 \\ 2\frac{\tilde{a}^{1/2}}{e}\frac{\sqrt{1 - e^2}\sin E}{\sqrt{1 - e^2 \cos^2 E}} \end{bmatrix}, \tag{8}$$

$$\frac{dE}{d\tilde{t}} = \frac{\tilde{a}^{-3/2}}{1 - e\cos E}\left(1 - \frac{2\tilde{a}^2}{e}\frac{(1 - e\cos E)\sin E}{\sqrt{1 - e^2 \cos^2 E}}\varepsilon\right).$$

The change of independent variable from time to eccentric anomaly leads to:

$$\frac{d\boldsymbol{\alpha}}{dE} = \frac{d\boldsymbol{\alpha}}{d\tilde{t}}\left(\frac{dE}{d\tilde{t}}\right)^{-1} = \frac{1-e\cos E}{1-\beta(E)\varepsilon}\begin{bmatrix} 2\tilde{a}^3\sqrt{\frac{1+e\cos E}{1-e\cos E}}\varepsilon \\ 2\tilde{a}^2\frac{(1-e^2)\cos E}{\sqrt{1-e^2\cos^2 E}}\varepsilon \\ 0 \\ 0 \\ 2\tilde{a}^2\frac{\sqrt{1-e^2}\sin E}{\sqrt{1-e^2\cos^2 E}}\varepsilon \\ \tilde{a}^{3/2} \end{bmatrix} = \boldsymbol{f}(E,\boldsymbol{\alpha},\varepsilon), \tag{9}$$

with $\beta(E) = \frac{2\tilde{a}^2}{e}\frac{(1-e\cos E)\sin E}{\sqrt{1-e^2\cos^2 E}}$.

To find an approximate solution to the system in Eq. (9), we apply a two-stage linearization procedure, first with respect to $\varepsilon$ (see Section 2.2), then with respect to $\boldsymbol{\alpha}$. The first linearization yields an equation in the form of Eq. (5), that is:

$$\frac{d\boldsymbol{\alpha}}{dE} \approx \boldsymbol{f}^0(E,\boldsymbol{\alpha}) + \boldsymbol{f}^1(E,\boldsymbol{\alpha})\varepsilon, \tag{10}$$

Linearizing further Eq. (10) with respect to the state vector about the chief orbital elements $\boldsymbol{\alpha}_c$, yields:

$$\frac{d\boldsymbol{\alpha}}{dE} \approx \boldsymbol{f}^0(E,\boldsymbol{\alpha}_c) + \boldsymbol{f}^1(E,\boldsymbol{\alpha}_c)\varepsilon + \nabla[\boldsymbol{f}^0(E,\boldsymbol{\alpha}_c) + \boldsymbol{f}^1(E,\boldsymbol{\alpha}_c)\varepsilon]\delta\boldsymbol{\alpha}, \tag{11}$$

where $\delta\boldsymbol{\alpha} = \boldsymbol{\alpha} - \boldsymbol{\alpha}_c$, represents the state perturbation and $\boldsymbol{A}(E,\varepsilon) = \nabla[\boldsymbol{f}^0(E,\boldsymbol{\alpha}_c) + \boldsymbol{f}^1(E,\boldsymbol{\alpha}_c)\varepsilon] = \boldsymbol{A}^0 + \boldsymbol{A}^1\varepsilon$ is the gradient with respect to $\boldsymbol{\alpha}$. An equivalent, more compact version of Eq (11) can be derived for the dynamics of the state perturbation as

$$\frac{d\delta\boldsymbol{\alpha}}{dE} \approx \boldsymbol{f}^0(E,\boldsymbol{\alpha}_c) + \boldsymbol{f}^1(E,\boldsymbol{\alpha}_c)\varepsilon + \boldsymbol{A}(E,\varepsilon)\delta\boldsymbol{\alpha} - \frac{d\boldsymbol{\alpha}_c}{dE} = \boldsymbol{f}^1(E,\boldsymbol{\alpha}_c)\varepsilon + \boldsymbol{A}(E,\varepsilon)\delta\boldsymbol{\alpha}, \tag{12}$$

where we have used: $\frac{d\boldsymbol{\alpha}_c}{dE} = \boldsymbol{f}^0(E,\boldsymbol{\alpha}_c)$. Explicit forms of $\boldsymbol{f}^0(\boldsymbol{\alpha}_c,E)$, $\boldsymbol{f}^1(\boldsymbol{\alpha}_c,E)$ and $\boldsymbol{A}(E)$ are given in Appendix A.

Eq.(12) is a system of time-variant non-homogeneous linear differential equations. A general solution can be expressed in terms of a state-transition matrix $\Phi$, as in:

$$\delta\boldsymbol{\alpha}(E_1) = \Phi(E_1,0)\left[\delta\boldsymbol{\alpha}(0) + \varepsilon\int_0^{E_1}\Phi^{-1}(E,0)\boldsymbol{f}^1(\boldsymbol{\alpha}_c,E)dE\right]. \tag{13}$$

The state transition matrix is by itself the solution of a matrix differential equation which can be integrated numerically, an approach that, however, would undermine the quest for an analytic solution. Here, instead, we seek for an approximate solution to the linear system making use of averaging theory. To this end, we will exploit the typical decoupling in the orbital elements' dynamics between the "fast" variable $\delta\tilde{t}$, the only one whose derivative features non-periodic components independent on $\varepsilon$ and the remaining, "slow" orbital elements $\delta\boldsymbol{\alpha}_r$, whose variation is periodic and proportional to $\varepsilon$. The two dynamical systems then read:

$$\frac{d\delta\boldsymbol{\alpha}_r}{dE} = \varepsilon \boldsymbol{f}_r^1(E,\boldsymbol{\alpha}_c) + \varepsilon \boldsymbol{A}_r^1(E,\boldsymbol{\alpha}_c)\delta\boldsymbol{\alpha},$$

$$\frac{d\delta\tilde{t}}{dE} = \varepsilon f_6^1(E,\boldsymbol{\alpha}_c) + \left(\boldsymbol{A}_6^0(E,\boldsymbol{\alpha}_c) + \varepsilon \boldsymbol{A}_6^1(E,\boldsymbol{\alpha}_c)\right)\delta\boldsymbol{\alpha}. \tag{14}$$

The first dynamical system is decoupled from the second because the last column of $\boldsymbol{A}_r^1$ is null, which is always true unless the perturbing acceleration depends explicitly on time. The solution method developed in the following, which will be tailored to the constant tangential acceleration model Eq. (9), may indeed be apply to any small perturbation force fulfilling such requirement.

## 2.4 Averaging of periodic systems

We are concerned with equations in standard form for averaging of the kind:

$$\dot{\boldsymbol{x}} = \epsilon g^1(\boldsymbol{x},t) = \epsilon(g(t) + G(t)\boldsymbol{x}). \tag{15}$$

Assuming that $g^1(\boldsymbol{x},t)$ is periodic in $t$, the solution can be tackled applying the so-called *improved first approximation* (see Sanders (Sanders, et al., 2007) p. 36), i.e. replacing the original differential equation with:

$$\dot{\boldsymbol{x}} \approx \epsilon g^1(\boldsymbol{z},t), \tag{16}$$

where $\boldsymbol{z}$ is the solution to the *averaged* system:

$$\dot{\boldsymbol{z}} = \epsilon \bar{g}^1(\boldsymbol{z}), \tag{17}$$

and

$$\bar{g}^1(\boldsymbol{x}) = \frac{1}{T}\int_0^T g^1(\boldsymbol{x},\tau)d\tau = \bar{g} + \bar{G}\boldsymbol{x}, \tag{18}$$

is the average of $g^1$, with the integration performed keeping $\boldsymbol{x}$ constant. Provided that the average system can be solved analytically, the approximate solution can then be expressed as the sum:

$$\boldsymbol{x}(t) \approx \boldsymbol{z}(t) + \boldsymbol{x}^{sp}(t), \tag{19}$$

where $\boldsymbol{x}^{sp}$, often referred to as short period term, is the solution to:

$$\dot{\boldsymbol{x}}^{sp} = \epsilon g^1(\boldsymbol{z},t) - \epsilon \bar{g}^1(\boldsymbol{z}), \tag{20}$$

which is assumed to fulfill the null-average condition $\frac{1}{T}\int_0^T \boldsymbol{x}^{sp}dt = \boldsymbol{0}$. By letting

$$d\boldsymbol{x}^{sp} = \int \dot{\boldsymbol{x}}^{sp} dt,$$

$$d\overline{\boldsymbol{x}^{sp}} = \frac{1}{T}\int_0^T d\boldsymbol{x}^{sp} dt, \tag{21}$$

then the short period contribution can be written as:

$$\boldsymbol{x}^{sp} = d\boldsymbol{x}^{sp} - d\overline{\boldsymbol{x}^{sp}}. \tag{22}$$

The average system is a set of linear, non-autonomous, differential equations with constant coefficients and constant forcing vector, whose solution is readily computed as:

$$\boldsymbol{z}(t) = e^{\epsilon t \bar{G}}\boldsymbol{z}_0 + \left[e^{\epsilon t \bar{G}} - I\right]\bar{G}^{-1}\bar{g}. \tag{23}$$

The formal solution to the short period differential equation is instead:

$$\boldsymbol{x}^{sp} = \epsilon(\mathcal{g}(t) - \bar{g}t) + \epsilon \int_0^T (G(t) - \bar{G})\boldsymbol{z}dt - d\overline{\boldsymbol{x}^{sp}}. \tag{24}$$

The integral at the right-hand side is in general more difficult to solve in analytical form. However, in the assumption that $\boldsymbol{x}_0 = \boldsymbol{0}$, which is the case for this work, $\boldsymbol{z}_0 = -\boldsymbol{x}^{sp}(\boldsymbol{0})$ has to be of order $\epsilon$. By inspection of Eq.(23), it follows that $\boldsymbol{z}$ is of order $\epsilon$ within a time interval $0 < t < \bar{t}/\epsilon$, where $\bar{t}$ is a constant independent of epsilon. One could then seek an approximation to the short period which retains only terms of order $\epsilon$, that is:

$$\boldsymbol{x}^{sp} = \epsilon(\mathcal{g}(t) - \bar{g}t) - d\overline{\boldsymbol{x}^{sp}}. \tag{25}$$

The final approximate solution becomes:

$$\boldsymbol{x} = \boldsymbol{z} + \epsilon(\mathcal{g}(t) - \bar{g}t) - d\overline{\boldsymbol{x}^{sp}}. \tag{26}$$

*2.5 Heuristic averaging of non-periodic systems*

If the system $g^1$ is non-periodic the results of 2.4 are in general not applicable, however we can heuristically retain the approximation $\dot{\boldsymbol{x}} \approx \epsilon g^1(\boldsymbol{z}, t)$ for integrating Eq. (15) without splitting between average and periodic contributions. To this end, we integrate Eq. (15) from the initial condition $\boldsymbol{x}_0$ as in:

$$\boldsymbol{x} \approx \boldsymbol{x}_0 + \epsilon \int (g(t) + G(t)\boldsymbol{z})dt = \boldsymbol{x}_0 + \epsilon \mathcal{g}(t) + \epsilon \int G(t)\boldsymbol{z}dt. \tag{27}$$

Since, in practice, the rightmost term is difficult to evaluate analytically because we have to consider $\boldsymbol{z}$ dependency with time, an alternative form will be used here. Using integration by parts yields to:

$$\boldsymbol{x} \approx \boldsymbol{x}_0 + \epsilon \mathcal{g}(t) + \epsilon \mathcal{G}(t)\boldsymbol{z} - \epsilon \int \mathcal{G}(t)\dot{\boldsymbol{z}}dt. \tag{28}$$

Assuming again $x_0 = 0$, so that $z$ is of order $\epsilon$, the last integral in Eq. (28) is at most of order $\epsilon^2$. Upon substitution of Eq (18) in place of $\dot{z}$, the leading term of the integral under study is:

$$\epsilon \int \mathcal{G}(t)\dot{z}dt = \epsilon^2 \int \mathcal{G}(t)dt\, \bar{g} + O(\epsilon^3). \tag{29}$$

Eventually, the approximation that will be used for solving the time equation is of the kind:

$$x \approx x_0 + \epsilon g(t) + \epsilon \mathcal{G}(t)z - \epsilon^2 \int \mathcal{G}(t)dt\, \bar{g}. \tag{30}$$

Note that the $\epsilon^2$ order only pertains to the approximation of the integral in Eq. (28), not to the final solution $x$, whose error depends on $z$ which is by itself an approximation of order $\epsilon$.

## 3 Approximate GPE solution

### 3.1 Solution of slow orbital elements

Application of Section 2.4 to the evolution of the slow orbital elements requires the solution of:

$$\frac{d\delta\bar{\alpha}_r}{dE} = \varepsilon\left(\bar{f}_r^1(\alpha_c) + B_r^1 \delta\bar{\alpha}_r\right),$$

$$\frac{d\delta\alpha_r^{sp}}{dE} = \varepsilon\left(f_r^1(E) - \bar{f}_r^1\right), \tag{31}$$

$$\text{with} \quad \delta\bar{\alpha}_{r0} = -\delta\alpha_r^{sp}(0),$$

The average function and Jacobian $\bar{f}_r^1$ and $B_r^1$ can be obtained by evaluating the primitives given in Appendix B between 0 and $2\pi$, i.e. following Eq.(18).

Being $B_r^1$ matrix in general non-invertible, the first of Eq. (31) is to be solved using Jordan-Block matrix decomposition $B_r^1 = UDU^{-1}$, where $U$ is a non-singular matrix and $D$ a block-diagonal matrix in the form:

$$D = \begin{bmatrix} 0 & 0 \\ 0 & d \end{bmatrix} \tag{32}$$

Then, using $e^{UDU^{-1}} = Ue^D U^{-1}$, the solution to the first of Eq. (31) is given by

$$\delta\bar{\alpha}_r = e^{B_r^1 \varepsilon E}\left(\delta\bar{\alpha}_{r0} + \varepsilon\, U \int_0^E e^{-D\varepsilon E'} dE'\, U^{-1}\, \bar{f}_r^1\right) \tag{33}$$

Leading to:

$$\delta\bar{\alpha}_r = e^{B_r^1 \varepsilon E}\left(\delta\bar{\alpha}_{r0} + \varepsilon U \begin{bmatrix} EI & 0 \\ 0 & (\varepsilon d)^{-1}(I - e^{-d\,\varepsilon E}) \end{bmatrix} U^{-1}\, \bar{f}_r^1\right) \tag{34}$$

Application of Eq. (21) and (22) to the second of Eq (31), leads to the short period component of the solution:

$$\delta \boldsymbol{\alpha}_r^{sp} = \varepsilon\big(\boldsymbol{f}_r^1(E) - \bar{\boldsymbol{f}}_r^1 E\big) - \frac{\varepsilon}{2\pi}\int_0^{2\pi}\big(\boldsymbol{f}_r^1(s) - \bar{\boldsymbol{f}}_r^1 s\big)ds. \tag{35}$$

Explicit expressions for $\boldsymbol{f}_r^1$ are given in Appendix B. The rightmost term in Eq.(35) requires integration of elliptic integrals and is therefore not solvable in closed form. The outcome is, however, a constant that has to be computed only once at the beginning of propagation. Although analytic approximations can be obtained through series expansion, in this work numerical integration was preferred to avoid truncation errors.

*3.2  Solution of time orbital element*

For the solution of the time equation, we make use of the improved first approximation $\dot{x} \approx \epsilon f^1(z,t)$, as in Section 2.5, retaining terms up to order $\epsilon^2$. However, due to the presence of terms of order 0 in the gradient, $\boldsymbol{A}^0$, Eq.(12) needs first to be reduced into the standard form for averaging of Eq.(15). Straight application of the variation of constants method (Sanders, et al., 2007) leads to the following standard system for the auxiliary variable $\boldsymbol{\varsigma} = \big(I + \mathcal{A}^0(E)\big)^{-1}\delta\boldsymbol{\alpha}$:

$$\dot{\boldsymbol{\varsigma}} = \varepsilon\big(I + \mathcal{A}^0(E)\big)^{-1}\big(\boldsymbol{f}_\varepsilon(E) + \boldsymbol{A}^1(E)\big(I + \mathcal{A}^0(E)\big)\boldsymbol{\varsigma}\big). \tag{36}$$

Thanks to the shape of $\mathcal{A}^0$ (see Eq. (B7) in Appendix B), $\big(I + \mathcal{A}^0(E)\big)^{-1} = I - \mathcal{A}^0(E)$. Moreover, for perturbations which do not depend explicitly on time, as it is the case here, the last column of $\boldsymbol{A}^1$ is null so that $\boldsymbol{A}^1(E)\mathcal{A}^0(E) = \boldsymbol{0}$, and the above simplifies to:

$$\dot{\boldsymbol{\varsigma}} = \varepsilon\big(I - \mathcal{A}^0(E)\big)\big(\boldsymbol{f}^1(E) + \boldsymbol{A}^1(E)\boldsymbol{\varsigma}\big) \tag{37}$$

We are interested here into the last component of $\delta\boldsymbol{\alpha}$, $\delta\tilde{t} = [\mathcal{A}_{61}^0 \quad \mathcal{A}_{62}^0 \quad ... \quad 1]\boldsymbol{\varsigma}$ and we need to solve:

$$\dot{\varsigma}_6 = \varepsilon[-\mathcal{A}_{61}^0 \quad -\mathcal{A}_{62}^0 \quad ... \quad 1]\big(\boldsymbol{f}^1(E) + \boldsymbol{A}^1(E)\boldsymbol{\varsigma}\big) = \varepsilon h^0(E) + \varepsilon \boldsymbol{h}^1(E)\boldsymbol{\varsigma}, \tag{38}$$

where:

$$h^0 = -\mathcal{A}_{61}^0 f_1^1(E) - \mathcal{A}_{62}^0 f_2^1(E) + f_6^1(E), \tag{39}$$

whereas $\boldsymbol{h}^1$ for a constant tangential thrust is a row vector with only the first two components different from 0:

$$\boldsymbol{h}^1 = \begin{bmatrix} -\mathcal{A}_{61}^0 A_{11}^1(E) - \mathcal{A}_{62}^0 A_{21}^1 + A_{61}^1(E) \\ -\mathcal{A}_{61}^0 A_{12}^1(E) - \mathcal{A}_{62}^0 A_{22}^1 + A_{62}^1(E) \\ 0 \\ 0 \\ 0 \\ 0 \end{bmatrix}^T. \tag{40}$$

Following 2.5, we use the approximation $\dot{x} \approx \epsilon f^1(z,t)$ where $z$ is the average solution $\bar{\boldsymbol{\varsigma}}$, to obtain:

$$\varsigma_6 = \varepsilon\left(\int h^0 dE + \int \boldsymbol{h}^1(E)\bar{\boldsymbol{\varsigma}}dE\right) + const \tag{41}$$

For the time equation, we retain up to the second order in $\varepsilon$ integrating by parts the right-hand side, yielding to:

$$\varsigma_6 = \varepsilon\left(\int h^0 dE + (\hbar^1(E) + c_{\hbar^1})\bar{\varsigma}\right) - \varepsilon^2 \int (\hbar^1(E) + c_{\hbar^1}) dE\, \bar{f}^1 + const, \quad (42)$$

where $c_{\hbar^1}$ is the integration constant arising from $\hbar^1$ indefinite integral. The integrals $\hbar^0$ and $\hbar^1$ can be analytically solved (although not in closed form) and are reported in Appendix C. The solution for the time equation can be eventually stated as:

$$\delta\tilde{t} = \mathcal{A}^0_{61}\delta a + \mathcal{A}^0_{62}\delta e + \varepsilon\left(\int h^0 dE + (\hbar^1(E) + c_{\hbar^1})\begin{bmatrix}\delta\bar{a}\\\delta\bar{e}\end{bmatrix}\right) - \varepsilon^2 \int (\hbar^1(E) + c_{\hbar^1}) dE\, \bar{f}^1 + const. \quad (43)$$

In Eq. (43) we have implicitly restricted the $\boldsymbol{h}^1$ vector to its not null elements and used the fact that $\delta\bar{\boldsymbol{\alpha}}_r = \bar{\varsigma}_r$.

## 4   Numerical validation

To justify the approach envisaged herein, we compared numerical integration of Eq. (1) obtained using Matlab ode113 solver with relative and absolute tolerances set equal to $10^{-13}$ (assumed as the ground truth), with the numerical integration of Eq. (12), using the same solver and settings, and the proposed analytical solution, Eqs. (34), (35), (43). Note that the error from numerically integrating Eq. (12) represents a limit for the achievable accuracy of our method, which approximates the doubly-linearized dynamics using averaging theory.

We took as a test case a reference LEO orbit with semi-major axis 8500km, eccentricity 0.2, null argument of pericenter and null eccentric anomaly at initial time, then considered a propagation time of 5 orbits and a perturbative constant small acceleration of $10^{-7}$ km/s$^3$.

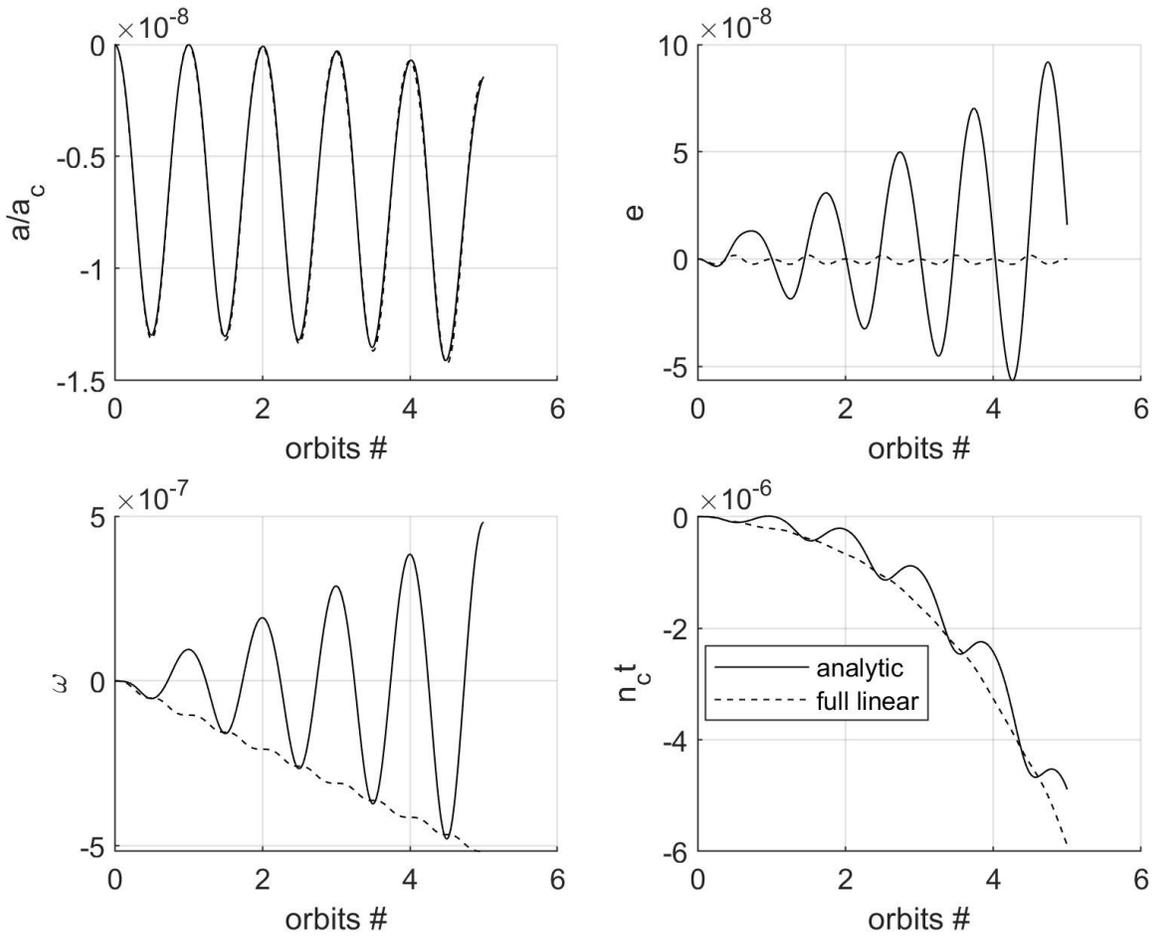

**Figure 1** Absolute errors of proposed solution and linearized solution with respect to the ground truth on the reference orbit

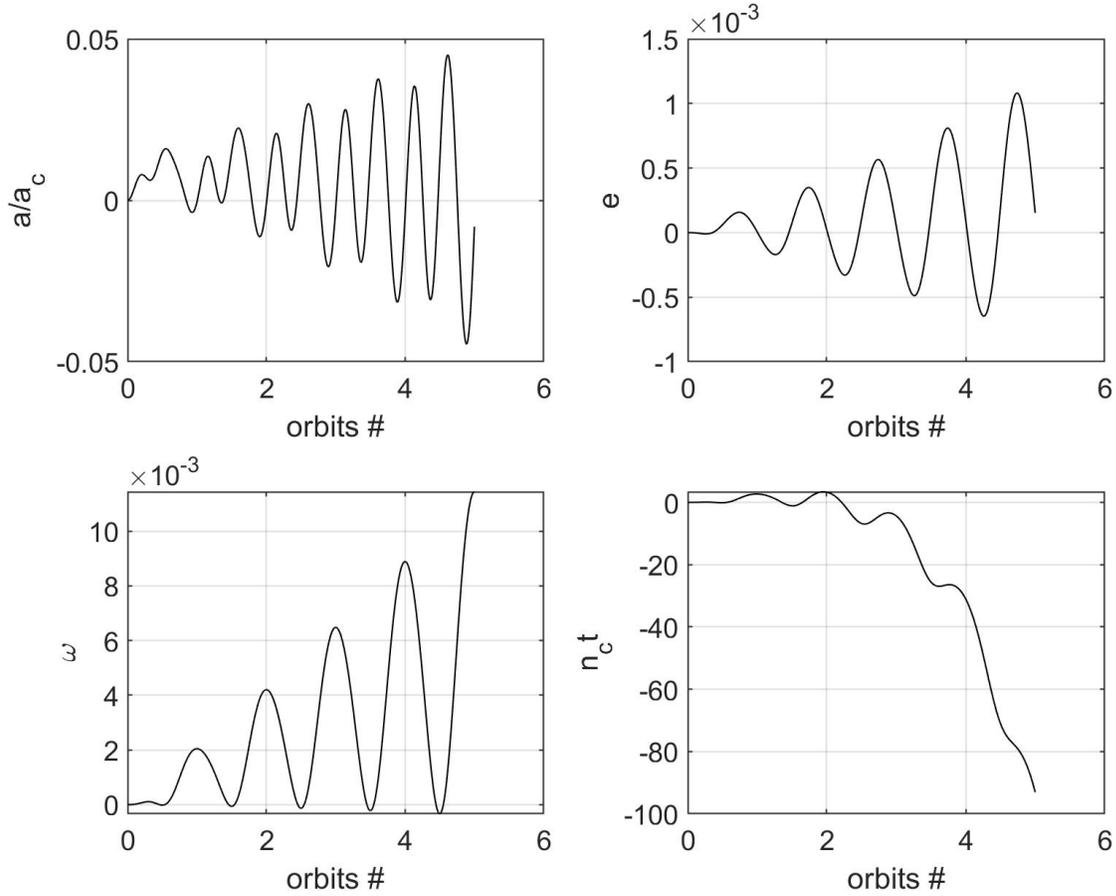

**Figure 2 Absolute errors of proposed solution with respect to the linear solution on the reference orbit but with perturbing acceleration of 1e-5 km/s$^3$**

Propagation errors are depicted in Figure 1. The analytic solution maintains the absolute errors in the order of $10^{-8}$ for the (dimensionless) semimajor axis, $10^{-7}$ for the eccentricity and argument of perigee, and $10^{-6}$ for the (dimensionless) time. In general, the semi-major axis and time variables are the ones tracking more closely the numerical integration of the linearized system.

For relatively large perturbation forces however (see Figure 2), the proposed solution, especially for the time element, drifts apart from the true linear solution. This is expected as the acceleration gets larger the root assumption of averaging theory degrades.

Table 1 reports the root mean square errors (non-dimensional) of the proposed averaged model with respect to the ground truth. Again, the performance degrades as the magnitude of the thrusting force increases. However, now the error is the joint result of the approximations introduced by the double stage linearization of GPE, plus those due to the subsequent application of averaging theory.

Table 1: Errors with respect to the ground truth with varying tangential acceleration (non-dimensional)

| | Input parameters | | | |
|---|---|---|---|---|
| $\varepsilon$ | $a/a_c$ | $e$ | $\omega$ | $n_c t$ |
| 1.81e-6 | 7.64e-11 | 3.50e-10 | 2.10e-09 | 2.24e-08 |
| 1.81e-5 | 7.93e-09 | 3.51e-08 | 2.10e-07 | 2.14e-06 |
| 1.81e-4 | 1.18e-06 | 3.56e-06 | 2.14e-05 | 2.45e-4 |
| 1.81e-3 | 8.48e-04 | 4.23e-04 | 2.62e-03 | 0.0634 |

Next, the impact of variations in eccentricity and the semi-major axis on the prediction error have been explored. Starting from the reference scenario of Figure 1, values of $a \in (6800, 8500, 10200)$ and $e \in (0.2, 0.5, 0.8)$ were tested, with results depicted respectively in Figure 3 and Figure 4. Notice that these orbits have been chosen solely for evaluation purposes and some of them are not physically meaningful, e.g. orbit with 8500km semi-major axis and 0.8 eccentricity.

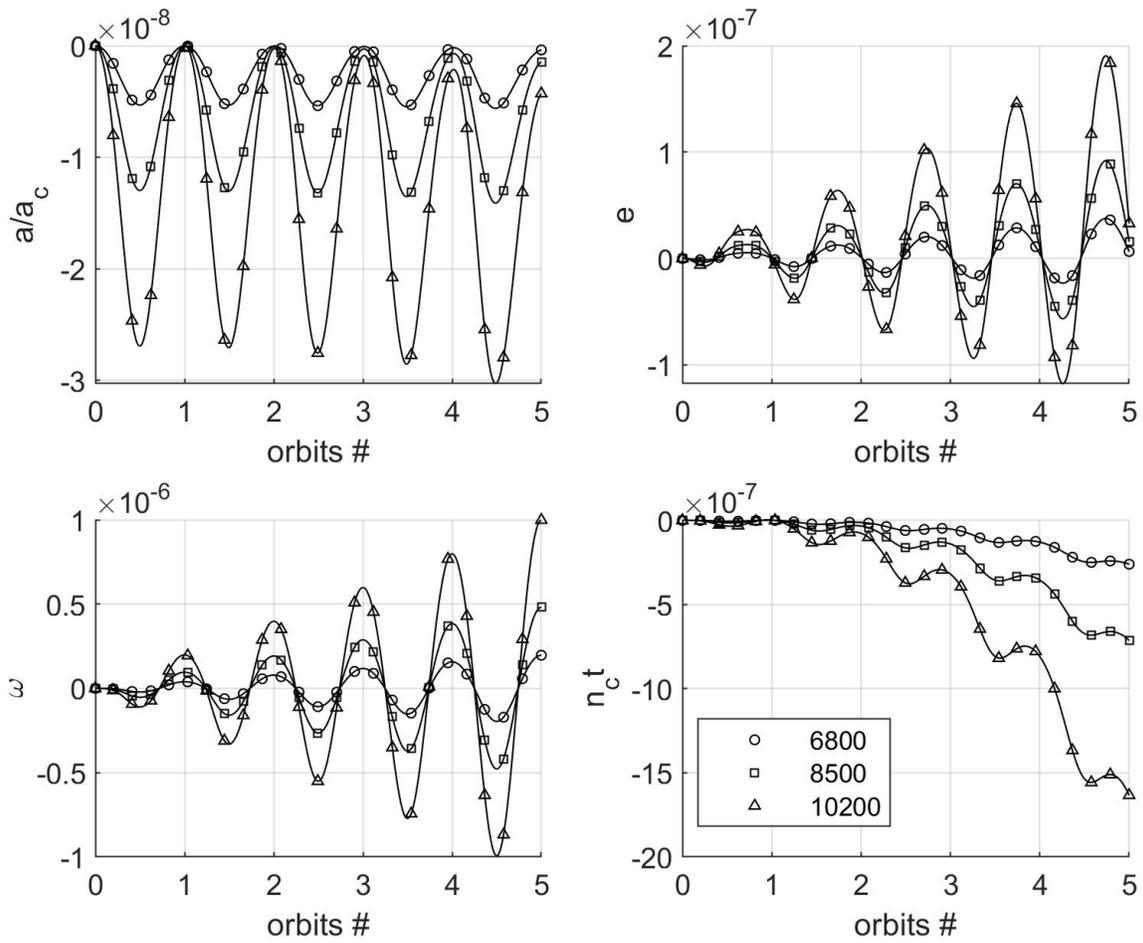

**Figure 3 Absolute errors of proposed solution with respect to the ground truth on the reference orbit changing semi-major axis**

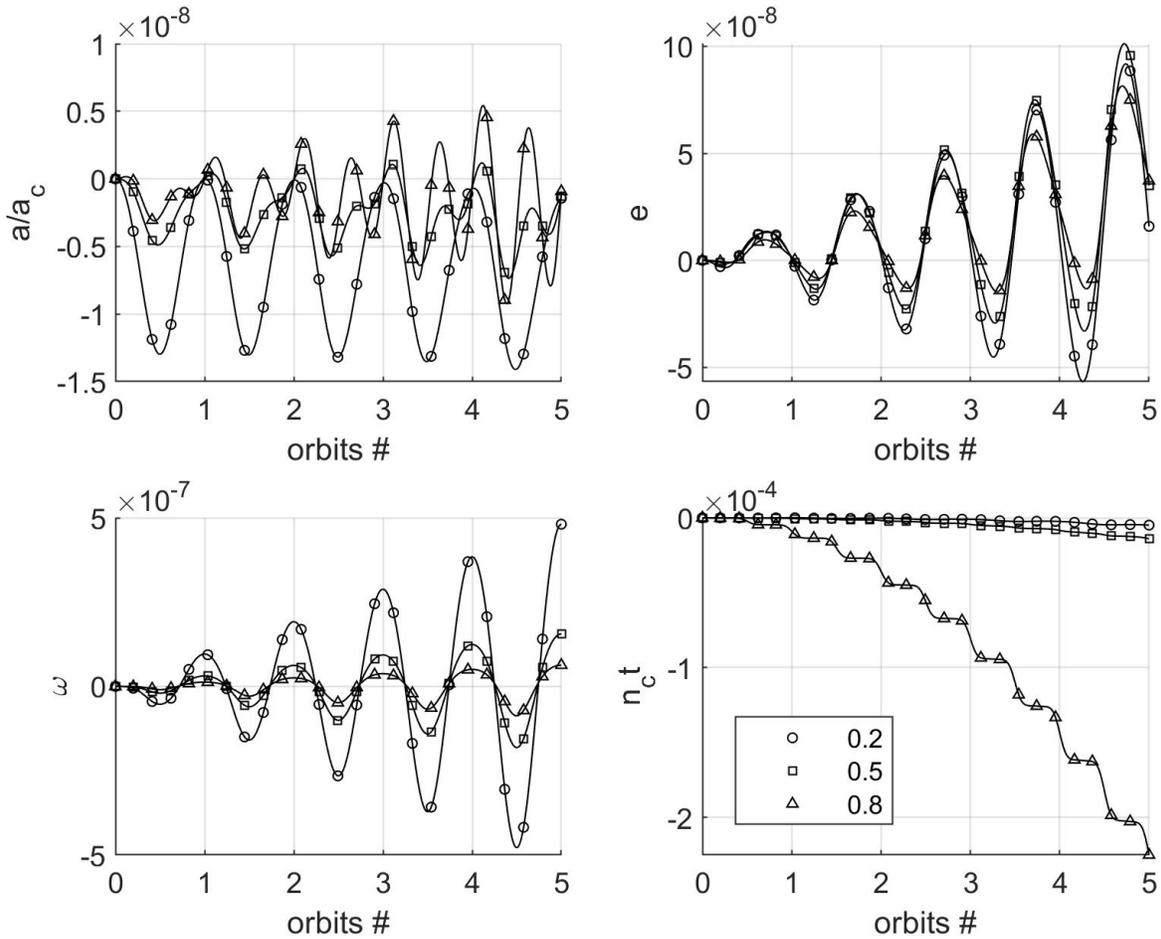

**Figure 4 Absolute errors of proposed solution with respect to the ground truth on the reference orbit changing eccentricity**

Changing the semi-major axis changes the orbital period and the error scales accordingly and almost proportionally. For this reason, the proposed solution seems to be tolerant to changes in semi-major axis.

As depicted in Figure 4, the error trends are only slightly affected by eccentricity within a certain range, except for the time element. Its solution, in fact, involves a few integrals which are approximated through Taylor series expansion around $e = 0$ see Appendix C, explaining the degraded accuracy obtained for the test case $e = 0.8$. Furthermore, increased errors are expected for very small eccentricity: since we have used singular orbital elements, the prediction eventually becomes less accurate as the eccentricity approaches zero, whereby the gradient becomes numerically very large amplifying errors in the orbital elements.

*4.1    Comparison with state of the art solutions*

We compare the prediction obtained using our model with those from the methods described in Gonzalo et al. (Gonzalo & Colombo, 2021) and Bombardelli (Bombardelli, et al., 2011), for the same test case of Figure 1. Figure 5 shows that the model proposed herein can accurately track the average trend of semi-major axis, eccentricity and time element. The largest improvements are obtained in the semi-major axes and time elements, whose errors are reduced respectively by almost two orders and by one order of magnitude. The argument of pericenter instead has no improvement and its evolution is exactly the same as the one proposed by Gonzalo et al, see Eq. (B1). This had to be expected as the average solution has null gradient for $\omega$ leaving therefore only the short period solution at first order, which is indeed identical to the one proposed in (Gonzalo & Colombo, 2021). Overall the *rms* value of the position error magnitude along the entire propagation interval reduces from 183.9 and 116.5 meters obtained respectively from Gonzalo et al. (Gonzalo & Colombo, 2021) and Bombardelli (Bombardelli, et al., 2011) down to 17.5 meters. The largest error being on the tangential component which still reduces from 139.3 and 84.3 to 12.6 meters respectively.

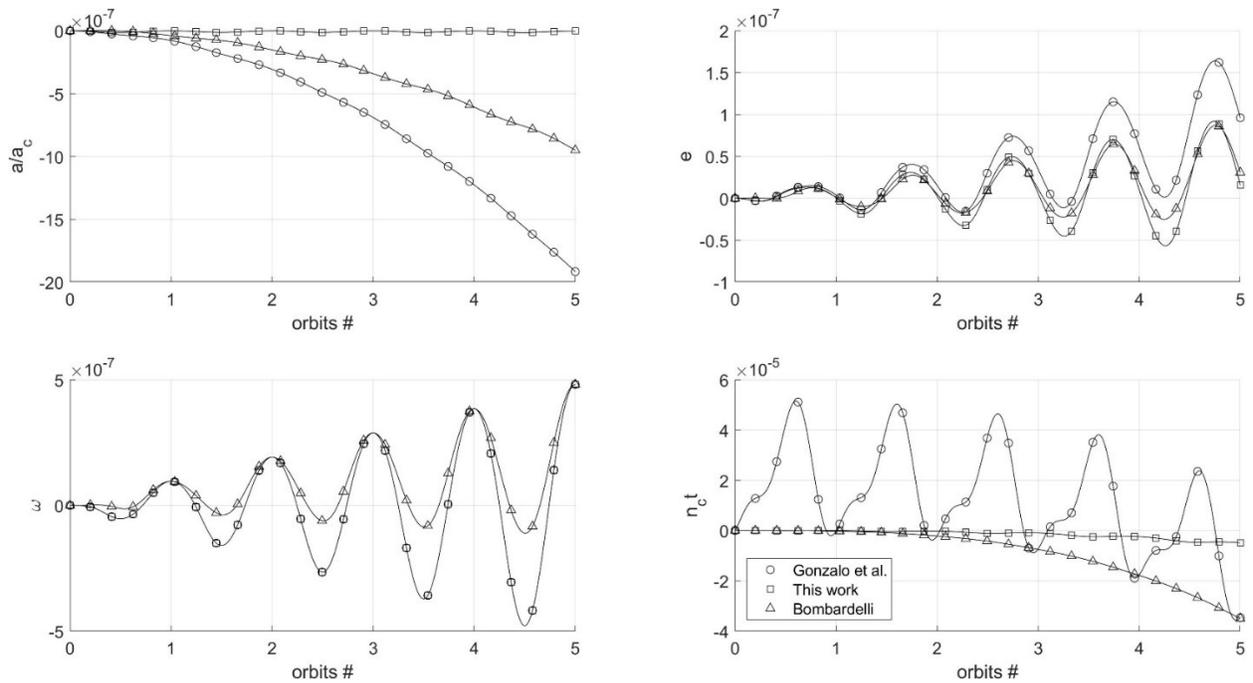

**Figure 5 Comparison between state of the art method and our proposed method on the reference orbit**

## 5    Conclusion

In this work, we presented an application of averaging techniques to obtain an approximated analytical solution to the problem of spacecraft motion subjected to a small perturbing acceleration. Although a tangentiality assumption was used for the showcase and comparison, the method envisaged herein is applicable to any small 3-dimensional thrust, as long as it is analytically integrable when placed into a linearized GPE model.

Results show that the obtained solution compares favorably with state of the art methods. The largest error is on the time element, ranging between 1 millisecond to 0.1 second for 5-orbit propagation and 1e-7 km/s$^3$ perturbation across the assumed ranges of semi-major axes and eccentricities. As an order of magnitude, this corresponds in LEO to an error in the tangential direction between 7m and 700m.

The error estimates strongly depend on the magnitude of the perturbing acceleration: an increase on the thrust magnitude of 1 order of magnitude increases the prediction error by approximately 2 orders of magnitude.

As a natural extension of the work presented herein, we aim at including other kinds of accelerations, most notably J2 perturbation as it is very significant in LEO. Further, we will consider the application of this method to a set of non-singular orbital elements, to enhance its flexibility.

## Appendix A. Gradients of GPE under tangential acceleration

Linearization of Eq. (9) about $\varepsilon = 0$ as in $f(\alpha, E, \varepsilon) \cong f^0(\alpha, E) + f^1(\alpha, E)\varepsilon$ yields to:

$$f^0(\alpha, E) = (1 - e\cos E) \begin{bmatrix} 0 \\ 0 \\ 0 \\ 0 \\ 0 \\ \tilde{a}^{3/2} \end{bmatrix}, \tag{A1}$$

$$f^1(\alpha, E) = (1 - e\cos E) \begin{bmatrix} 2\tilde{a}^3 \sqrt{\dfrac{1 + e\cos E}{1 - e\cos E}} \\ 2\tilde{a}^2 \dfrac{(1 - e^2)\cos E}{\sqrt{1 - e^2 \cos^2 E}} \\ 0 \\ 0 \\ 2\tilde{a}^2 \dfrac{\sqrt{1 - e^2}\sin E}{\sqrt{1 - e^2 \cos^2 E}} \\ \tilde{a}^{\frac{3}{2}}\beta(E) \end{bmatrix}. \tag{A2}$$

Recalling that $\beta(E) = \dfrac{2\tilde{a}^2}{e} \dfrac{(1-e\cos E)\sin E}{\sqrt{1-e^2\cos^2 E}}$.

Then, when evaluated at the chief orbital elements where $\tilde{a} = 1$, we have:

$$\boldsymbol{f}^0(\boldsymbol{\alpha}_c, E) = \begin{bmatrix} 0 \\ 0 \\ 0 \\ 0 \\ 0 \\ 1 - e_c \cos E \end{bmatrix}$$

$$\boldsymbol{f}^1(\boldsymbol{\alpha}_c, E) = \begin{bmatrix} 2\sqrt{1 - e_c^2 \cos^2 E} \\ 2(1 - e_c^2)\sqrt{\dfrac{1 - e_c \cos E}{1 + e_c \cos E}} \cos E \\ 0 \\ 0 \\ 2 \dfrac{\sqrt{1 - e_c^2}}{e_c} \sqrt{\dfrac{1 - e_c \cos E}{1 + e_c \cos E}} \sin E \\ \dfrac{2}{e_c}(1 - e_c \cos E) \sqrt{\dfrac{1 - e_c \cos E}{1 + e_c \cos E}} \sin E \end{bmatrix}. \quad (A3)$$

The Jacobian matrix $\boldsymbol{A}(E) = \boldsymbol{A}^0 + \boldsymbol{A}^1 \varepsilon$ has only the first two columns including non-zero elements:

$$\frac{\partial \boldsymbol{f}(\boldsymbol{\alpha}, E)}{\partial \tilde{a}} = \begin{bmatrix} 6\tilde{a}^2 \sqrt{1 - e^2 \cos^2 E}\, \varepsilon \\ 4\tilde{a}(1 - e^2)\sqrt{\dfrac{1 - e \cos E}{1 + e \cos E}} \cos E\, \varepsilon \\ 0 \\ 0 \\ 4\tilde{a}\dfrac{\sqrt{1 - e^2}}{e}\sqrt{\dfrac{1 - e \cos E}{1 + e \cos E}} \sin E\, \varepsilon \\ (1 - e \cos E)\left(\dfrac{3\tilde{a}^{1/2}}{2} + \dfrac{7\tilde{a}^{5/2}}{e}\sqrt{\dfrac{1 - e \cos E}{1 + e \cos E}} \sin E\, \varepsilon \right) \end{bmatrix}$$

$$\frac{\partial \boldsymbol{f}(\boldsymbol{\alpha}, E)}{\partial e} = \begin{bmatrix} -2\tilde{a}^3 \dfrac{e \cos^2 E}{\sqrt{1 - e^2 \cos^2 E}} \varepsilon \\ -2\tilde{a}^2 \sqrt{\dfrac{1 - e \cos E}{1 + e \cos E}} \cos E \left[2e + (1 - e^2)\dfrac{\cos E}{1 - e^2 \cos^2 E}\right] \varepsilon \\ 0 \\ 0 \\ -\dfrac{2\tilde{a}^2}{e\sqrt{1 - e^2}}\sqrt{\dfrac{1 - e \cos E}{1 + e \cos E}} \sin E \left[\dfrac{1}{e} + \dfrac{\cos E}{1 - e^2 \cos^2 E}\right] \varepsilon \\ -\tilde{a}^{3/2} \cos E - \dfrac{2\tilde{a}^{7/2}}{e^2}\sqrt{\dfrac{1 - e \cos E}{1 + e \cos E}} \dfrac{1 + 2e \cos E}{1 + e \cos E} \sin E\, \varepsilon \end{bmatrix}. \quad (A4)$$

Evaluation at the chief orbit follows directly upon substitution of $\tilde{a} = 1$ and $e = e_c$.

**Appendix B. Integral of GPE and its gradient under tangential acceleration**

In the following, $F$, $\mathcal{E}$, and $\Pi$ will be used to denote the incomplete elliptic integrals of the first, second and third kind respectively. The analytic primitive of $\boldsymbol{f}^1(\boldsymbol{\alpha}_c, E)$ indicated in the text as $\boldsymbol{f}^1(t) = [\delta\tilde{a}^I, \delta e^I, \delta i^I, \delta\Omega^I, \delta\omega^I, \delta\tilde{t}^I]^T$, read:

$$\delta\tilde{a}^I(E) = 2\sqrt{1 - e_c^2}\mathcal{E}[E, k],$$

$$\delta e^I(E) = 2\frac{1 - e_c^2}{e_c}\left\{\frac{1}{2}\ln\frac{\sqrt{1 - e_c^2\cos^2 E} + e_c\sin E}{\sqrt{1 - e_c^2\cos^2 E} - e_c\sin E} - \frac{1}{\sqrt{1 - e_c^2}}F[E, k] + \sqrt{1 - e_c^2}\mathcal{E}[E, k]\right\},$$

$$\delta\omega^I(E) = 2\frac{\sqrt{1 - e_c^2}}{e_c}\left(2\mathrm{asin}\sqrt{\frac{1 - e_c\cos E}{2}} - \sqrt{1 - e_c^2\cos^2 E}\right)\Bigg|_0^E, \quad (B1)$$

$$\delta i^I = 0, \delta\Omega^I = 0,$$

$$\delta\tilde{t}^I(E) = -\frac{1}{e_c^2}\left(\sqrt{\frac{1 + e_c\cos E}{1 - e_c\cos E}}(e_c^2\cos^2 E - 5e_c\cos E + 4) - 6\,\mathrm{asin}\sqrt{\frac{1 - e\cos E}{2}}\right)\Bigg|_0^E,$$

where $k = \frac{e_c^2}{(e_c^2 - 1)}$. Note that the first three equations were first obtained by Gonzalo et al. (Gonzalo, et al., 2019).

To illustrate the integrals of the gradient, start by defining:

$$\mathcal{A} = \left[\int\frac{\partial\boldsymbol{f}(\boldsymbol{\alpha}, E)}{\partial\tilde{a}}\Bigg|_{\boldsymbol{\alpha}_c}dE \quad \int\frac{\partial\boldsymbol{f}(\boldsymbol{\alpha}, E)}{\partial e}\Bigg|_{\boldsymbol{\alpha}_c}dE \quad \boldsymbol{0}_{6\times 4}\right]. \quad (B2)$$

The first column (semi-axis derivatives) is computed through:

$$\int\frac{\partial\boldsymbol{f}(\boldsymbol{\alpha}, E)}{\partial\tilde{a}}\Bigg|_{\boldsymbol{\alpha}_c}dE = \begin{bmatrix} 3\delta a^I(E)\varepsilon \\ 2\delta e^I(E)\varepsilon \\ 0 \\ 0 \\ 2\delta\omega^I(E)\varepsilon \\ \frac{3}{2}(E - e_c\sin E) + \frac{7}{2}\delta t^I(E)\varepsilon \end{bmatrix}. \quad (B3)$$

The elements of the second column (eccentricity derivatives) are:

$$\mathcal{A}_{12} = -2e_c\varepsilon \frac{(F[E|k] + (e_c^2 - 1)\mathcal{E}[E|k])}{e_c^2\sqrt{1-e_c^2}}$$

$$\mathcal{A}_{22} = -\frac{2e_c}{(1-e_c^2)}\delta e^I(E) - 2\varepsilon(1-e_c^2)\int \frac{\cos^2 E}{\sqrt{1-e_c^2\cos^2 E}\,(1+e_c\cos E)}dE$$

$$= -\frac{2e_c}{(1-e_c^2)}\delta e^I(E)$$

$$-\frac{2\varepsilon}{e_c^2\sqrt{(e_c-1)^2(1-e_c^2)}}\Bigg(2\,\Pi\left[\frac{2e_c}{e_c-1};\frac{H}{2}\Big|\frac{2m}{m-1}\right](e_c^4+1)(e_c+1)$$

$$+ 3F\left[\frac{H}{2}\Big|\frac{2m}{m-1}\right](e_c+1)(-e_c^3+e_c^2+e_c-1)$$

$$- \mathcal{E}\left[\frac{H}{2}\Big|\frac{2m}{m-1}\right](e_c+1)(e_c^3-3e_c^2+3e_c-1)\Bigg)$$

(B4)

$$\mathcal{A}_{32} = \mathcal{A}_{42} = 0$$

$$\mathcal{A}_{52} = -\frac{2\varepsilon}{e_c\sqrt{1-e_c^2}}\left(\frac{2\mathrm{asin}\sqrt{\frac{1-e_c\cos E}{2}} - \sqrt{1-e_c^2\cos^2 E}}{e_c^2} - \frac{\mathrm{asin}(e_c\cos E) + \sqrt{\frac{1-e_c\cos E}{1+e_c\cos E}}}{e_c^2}\right)$$

$$\mathcal{A}_{62} = -\sin E + \frac{4\varepsilon}{e_c^3}\left(\sqrt{\frac{1-e_c\cos E}{1+e_c\cos E}}(2+e_c\cos E) + 3\,\mathrm{asin}\sqrt{\frac{1-e_c\cos E}{2}}\right).$$

The integral in the expression of $\mathcal{A}_{22}$ has been obtained using the following change of variable:

$$H = 2\,\mathrm{atan}\left(\sqrt{\frac{1-e_c}{1+e_c}}\tan\left(\frac{E}{2}\right)\right) \rightarrow \frac{dH}{dE} = \frac{\sqrt{1-e_c^2}}{1+e_c\cos E},$$

so that $\cos E = \frac{\cos H - e_c}{1 - e_c\cos H}$. Letting $m = \frac{2e_c}{1+e_c^2}$, the integral becomes:

$$\int \frac{\cos^2 E}{\sqrt{1-e_c^2\cos^2 E}\,(1+e_c\cos E)}dE = \frac{1}{\sqrt{1-e_c^2}}\int \frac{(\cos H - e_c)^2}{(1-e_c\cos H)\sqrt{(1-e_c^4)(1-m\cos H)}}dH \quad \text{(B5)}$$

The righthand side can then be split into the sum of the following three terms:

$$\frac{1}{\sqrt{1-e_c^4}} \int \frac{(1+e_c^2)/e_c^2}{(1-e_c \cos H)\sqrt{1-m \cos H}} dH = \frac{(1+e_c^2)/e_c^2}{\sqrt{1-e_c^4}} 2 \frac{\Pi\left[\frac{2e_c}{e_c-1}; \frac{H}{2} \middle| \frac{2m}{m-1}\right]}{(1-e_c)\sqrt{1-m}},$$

$$\frac{-1/e_c^2}{\sqrt{1-e_c^4}} \int \frac{1+e_c \cos H}{\sqrt{1-m \cos H}} dH =$$

$$= \frac{-\frac{1}{e_c^2}}{\sqrt{1-e_c^4}} \frac{2}{m\sqrt{1-m}} \left( (e_c+m)F\left[\frac{H}{2} \middle| \frac{2m}{m-1}\right] + e_c(m-1)\mathcal{E}\left[\frac{H}{2} \middle| \frac{2m}{m-1}\right] \right), \quad (B6)$$

$$\frac{1+e_c^2}{\sqrt{1-e_c^4}} \int \frac{\sqrt{1-m \cos H}}{1-e_c \cos H} dH =$$

$$= \frac{1+e_c^2}{\sqrt{1-e_c^4}} \frac{2}{e_c(e_c-1)\sqrt{1-m}} \left( m(e_c-1)F\left[\frac{H}{2} \middle| \frac{2m}{m-1}\right] \right.$$

$$\left. + (m-e_c)\Pi\left[\frac{2e_c}{e_c-1}; \frac{H}{2} \middle| \frac{2m}{m-1}\right] \right).$$

The Keplerian and perturbative contributions to the gradient integral can be split as in:

$$\mathcal{A} = \mathcal{A}^0 + \mathcal{A}^1 \varepsilon,$$

where:

$$\mathcal{A}^0 = \begin{bmatrix} \mathbf{0}_{5\times 2} & & \mathbf{0}_{5\times 4} \\ \frac{3}{2}(E - e_c \sin E) & -\sin E & \mathbf{0}_{1\times 4} \end{bmatrix}, \quad (B7)$$

and $\mathcal{A}^1 \varepsilon$ is the remaining part of the sum.

**Appendix C. Integral of linearized GPE time equation**

The solution of the time equation (43) requires computing integrals of $h^0$, $h^1$ and $\hbar^1$, which in turn involves integrals of both $\mathcal{A}^0$ and $\mathcal{A}^1$. Some of those could not be solved in closed form and the following McLaurin expansions at 4-th order have been used:

$$\sqrt{1-e_c^2 \cos^2 E} = 1 - \frac{e_c^2}{2}\cos^2 E - \frac{e_c^4}{8}\cos^4 E + O(e_c^6)$$

$$\frac{1}{\sqrt{1-e_c^2 \cos^2 E}} = 1 + \frac{e_c^2}{2}\cos^2 E + \frac{e_c^4}{8}\cos^4 E + O(e_c^6)$$

(C1)

Also, the following integrals are handy in deriving the final result:

$$I_{c2} = \int E \cos^2 E \, dE = \frac{E^2}{4} - \frac{\sin^2 E}{8} + \frac{\cos^2 E}{8} + \frac{E}{2} \sin E \cos E$$

$$I_{c4} = \int E \cos^4 E \, dE = \frac{3E^2}{16} + \frac{E \sin(2E)}{4} + \frac{E \sin(4E)}{32} + \frac{\cos(2E)}{8} + \frac{\cos(4E)}{128}$$

$$I_{c6} = \int E \cos^6 E \, dE$$

$$= \frac{5E^2}{32} + \frac{15E \sin(2E)}{64} + \frac{3E \sin(4E)}{64} + \frac{E \sin(6E)}{192} + \frac{15 \cos(2E)}{128} + \frac{3 \cos(4E)}{256}$$

$$+ \frac{\cos(6E)}{1152}$$

$$I_{sc} = \int \sin E \sqrt{\frac{1 - e \cos E}{1 + e \cos E}} \cos E \, dE \qquad (C2)$$

$$= \frac{-1}{2e_c^2} \left( \frac{-e_c^3 \cos^3 E + 2e_c^2 \cos^2 E + e_c \cos E - 2}{\sqrt{1 - e_c^2 \cos^2 E}} + 2 \, a\sin \sqrt{\frac{1 - e_c \cos E}{2}} \right)$$

$$I_{sc2} = \int \sin E \frac{\cos^2 E}{(1 + e \cos E)\sqrt{1 - e_c^2 \cos^2 E}} \, dE = \frac{\sqrt{1 - e_c^2 \cos^2 E} \, (2 + e_c \cos E)}{1 + e_c \cos E} + \mathrm{asin}(e_c \cos E)$$

$$I_{sc3} = \int \sin E \frac{\cos^2 E}{\sqrt{1 - e_c^2 \cos^2 E}} \, dE = \frac{\cos E}{2e_c^2} \sqrt{1 - e_c^2 \cos^2 E} - \frac{\mathrm{asin}(e_c \cos E)}{2e_c^3}$$

$$I_s = \int \sin E \sqrt{1 - e_c^2 \cos^2 E} \, dE = -\frac{\cos E}{2} \sqrt{1 - e_c^2 \cos^2 E} - \frac{\mathrm{asin}(e_c \cos E)}{2e_c},$$

where the last four integrals have been obtained with a symbolic manipulator.

We start from the integral of $h^0(E)$ in Eq. (43), that is:

$$\int h^0 dE = \int (-\mathcal{A}_{61}^0 f_1^1 - \mathcal{A}_{62}^0 f_2^1 + f_6^1) dE \qquad (C3)$$

The last integral has already been given in Appendix B, the second can be solved in closed form to give:

$$\int -\mathcal{A}_{62}^0 f_2^1 dE = (1 - e_c^2) \, 2I_{sc} \qquad (C4)$$

For the first integral, the expansion (C1) has to be used instead because of the linear part in $E$ appearing in $\mathcal{A}_{61}^0$.

$$\int -\mathcal{A}_{61}^0 f_1^1 dE \cong -\frac{3}{2}\left(E^2 - e_c^2 \, I_{c2} - \frac{e_c^4}{4} I_{c4}\right) + 3e_c I_s. \qquad (C5)$$

We now compute the integral $\boldsymbol{h}^1(E)$ from Eq. (40), that is a row vector $\boldsymbol{h}$ with two non-null components:

$$\boldsymbol{\hbar} = \int \boldsymbol{h}^1 dE = \begin{bmatrix} \int -\mathcal{A}^0{}_{61}A^1_{11} - \mathcal{A}^0{}_{62}A^1_{21} + A^1_{61} dE \\ \int -\mathcal{A}^0{}_{61}A^1_{12} - \mathcal{A}^0{}_{62}A^1_{22} + A^1_{62} dE \end{bmatrix}^T$$

The rightmost addends are given in Appendix B being respectively $\mathcal{A}^1_{61}$ and $\mathcal{A}^1_{62}$. The remaining four terms are approximated through the expansions adopted above:

$$\int -\mathcal{A}^0_{61}A^1_{11} dE \cong -9\left(\frac{E^2}{2} - \frac{e_c^2}{2}I_{c2} - \frac{e_c^4}{8}I_{c4}\right) + 9e_c I_s$$

$$\int -\mathcal{A}^0_{62}A^1_{21} dE = 4(1 - e_c^2)I_{sc}$$

$$\int -\mathcal{A}^0_{61}A^1_{12} dE \cong 3e_c\left(I_{c2} + \frac{e_c^2}{2}I_{c4} + \frac{e_c^4}{8}I_{c6}\right) - 3e_c^2 I_{sc3}$$

$$\int -\mathcal{A}^0_{62}A^1_{22} dE = -2(2e_c I_{sc} + (1 - e_c^2) I_{sc2})$$

(C6)

To evaluate $\int \boldsymbol{\hbar} dE$, we start by computing the integrals of $\mathcal{A}^1_{61}$ and $\mathcal{A}^1_{62}$. To this aim, the following integrals are useful:

$$I_{c0F} = \int \frac{1}{\sqrt{1 - e_c^2 \cos^2 E}} dE = \frac{F[E|k]}{\sqrt{1 - e_c^2}}$$

$$I_{c1F} = \int \frac{\cos E}{\sqrt{1 - e_c^2 \cos^2 E}} dE = \frac{1}{e_c} \operatorname{asinh}\left(\frac{e_c}{\sqrt{1 - e_c^2}} \sin E\right)$$

$$I_{c2F} = \int \frac{\cos^2 E}{\sqrt{1 - e_c^2 \cos^2 E}} dE = \frac{F[E|k] + (e_c^2 - 1)\mathcal{E}[E|k]}{e_c^2\sqrt{1 - e_c^2}}$$

$$I_{c3F} = \int \frac{\cos^3 E}{\sqrt{1 - e_c^2 \cos^2 E}} dE = \frac{(e_c^2 + 1)\operatorname{atanh}\left(\frac{e_c \sin E}{\sqrt{1 - e_c^2 \cos^2 E}}\right) - e_c \sin E \sqrt{1 - e_c^2 \cos^2 E}}{2e_c^3}$$

$$I_{EsF} = \int \frac{E \sin E}{\sqrt{1 - e_c^2 \cos^2 E}} dE \cong \int E \sin E \, dE + \frac{e_c^2}{2}\int E \sin E \cos^2 E \, dE + \frac{3e_c^4}{8}\int E \sin E \cos^4 E \, dE.$$

(C7)

For the second of the above we have used the integral 630.11 form (Byrd & Friedman, 1971), whereas $I_{c2F}$ and $I_{c3F}$ have been obtained with a symbolic manipulator and $I_{EsF}$ approximated using the expansion in Eq. (C1). The following recurrence is also handy in the derivation:

$$\int \sqrt{\frac{1 \pm e_c \cos E}{1 \mp e_c \cos E}} \cos^i E \, dE = I_{ciF} \pm e_c I_{c(i+1)F},$$

(C8)

that can be derived by multiplying and dividing by $\sqrt{1 \pm e_c \cos E}$ which is always different from 0 for elliptic orbits.

Then:

$$\int \mathcal{A}_{61}^1 \, dE = \frac{7}{2}\left[-\frac{4}{e_c^2}\int \sqrt{\frac{1+e_c\cos E}{1-e_c\cos E}}\,dE + \frac{5}{e_c}\int \sqrt{\frac{1+e_c\cos E}{1-e_c\cos E}}\cos E\,dE\right.$$

$$\left.-\int \sqrt{\frac{1+e_c\cos E}{1-e_c\cos E}}\cos^2 E\,dE + \frac{6}{e_c^2}\int \mathrm{asin}\sqrt{\frac{1-e_c\cos E}{2}}\,dE\right]$$

$$= \frac{7}{2}\left[-\frac{4}{e_c^2}(I_{c0F} + e_c\,I_{c1F}) + \frac{5}{e_c}(I_{c1F} + e_c\,I_{c2F})dE - (I_{c2F} + e_c\,I_{c3F})\right.$$

$$\left.+\frac{6}{e_c^2}\left(E\,\mathrm{asin}\sqrt{\frac{1-e_c\cos E}{2}} - \frac{e_c}{2}I_{EsF}\right)\right]$$

(C9)

$$\int \mathcal{A}_{62}^1 \, dE = \frac{4}{e_c^3}\int\left(\sqrt{\frac{1-e_c\cos E}{1+e_c\cos E}}(2+e_c\cos E) + 3\,\mathrm{asin}\sqrt{\frac{1-e_c\cos E}{2}}\right)dE$$

$$= \frac{4}{e_c^3}\left[2(I_{c0F} - e_c\,I_{c1F}) + e_c(I_{c1F} - e_c\,I_{c2F}) + 3\left(E\,\mathrm{asin}\sqrt{\frac{1-e_c\cos E}{2}} - \frac{e_c}{2}I_{EsF}\right)\right].$$

To complete the integrals in $\int \hbar dE$ we then need the following:

$$\int I_{c2}\,dE = \int\left(\frac{E^2}{4} - \frac{\sin^2 E}{8} + \frac{\cos^2 E}{8} + \frac{E}{2}\sin E\cos E\right)dE$$

$$\int I_{c4}\,dE = \int\left(\frac{3E^2}{16} + \frac{E\sin(2E)}{4} + \frac{E\sin(4E)}{32} + \frac{\cos(2E)}{8} + \frac{\cos(4E)}{128}\right)dE$$

$$\int I_{c6}\,dE = \int\left(\frac{5E^2}{32} + \frac{15E\sin(2E)}{64} + \frac{3E\sin(4E)}{64} + \frac{E\sin(6E)}{192} + \frac{15\cos(2E)}{128} + \frac{3\cos(4E)}{256}\right.$$

$$\left.+\frac{\cos(6E)}{1152}\right)dE$$

(C10)

$$\int I_{sc}\,dE = \frac{-1}{2e_c^2}\int\left(\frac{-e_c^3\cos^3 E + 2e_c^2\cos^2 E + e_c\cos E - 2}{\sqrt{1-e_c^2\cos^2 E}} + 2\,\mathrm{asin}\sqrt{\frac{1-e_c\cos E}{2}}\right)dE$$

$$\int I_{sc2}\,dE = \int \frac{1}{e_c^3}\left(\frac{\sqrt{1-e_c^2\cos^2 E}\,(2+e_c\cos E)}{1+e_c\cos E} + \mathrm{asin}(e_c\cos E)\right)dE$$

$$\int I_{sc3}\,dE = \int \frac{\cos E}{2e_c^2}\sqrt{1-e_c^2\cos^2 E} - \frac{\mathrm{asin}(e_c\cos E)}{2e_c^3}\,dE$$

$$\int I_s dE = \int -\frac{\cos E}{2}\sqrt{1-e_c^2\cos^2 E} - \frac{\text{asin}(e_c \cos E)}{2e_c} dE.$$

Most of those integrals can be expressed in terms of others already described above or computed by any symbolic calculator. The only new ones are:

$$\int \text{asin}(e_c \cos E)\, dE \cong E\, \text{asin}(e_c \cos E) + e_c I_{EsF}, \qquad (C11)$$

and

$$\int \cos E \sqrt{1-e_c^2 \cos^2 E}\, dE = \frac{\sin E}{2}\sqrt{1-e_c^2 \cos^2 E} + (1-e_c^2)\frac{\text{asinh}\left(\frac{e_c}{\sqrt{1-e_c^2}}\sin E\right)}{2e_c} \qquad (C12)$$

The former has been approximated through integration by parts and then using McLaurin expansion around $e_c = 0$ for the integrand functions. The latter is obtained through integration by parts and applying integral 630.11 form (Byrd & Friedman, 1971).